\documentstyle[11pt,newpasp,psfig]{article}

\begin{document}

\title{Order and Chaos in the Local Disc Stellar Kinematics}
\author{R. Fux}
\affil{Research School of Astronomy and Astrophysics,
       Mount Stromlo Observatory, Canberra, Australia}

\begin{abstract}
The effects of the Galactic bar on the velocity distribution of
old disc stars in the Solar neighbourhood are investigated using
high-resolution 2D test particle simulations. A~detailed orbital
analysis reveals that the structure of the $U-V$ distribution
in these simulations is closely related to the phase-space extent
of regular and chaotic orbits. At low angular momentum and for a
sufficiently strong bar, stars mainly follow chaotic orbits which
may cross the corotation radius and the $U-V$ contours follow
lines of constant Jacobi's integral except near the regions
occupied by weakly populated eccentric regular orbits. These
properties can naturally account for the observed outward motion
of the local high asymmetric drift stars.
\end{abstract}

\section{Introduction}

The observed local stellar velocity distribution in the $U-V$ plane, where $U$
is the radial velocity positive towards the anti-centre and $V$ the azimuthal
velocity positive towards galactic rotation, betrays a large stream of old
disc stars with an asymmetric drift $S\approx 40-50$~km\,s$^{-1}$ and
$\overline{U}>0$ (e.g. Dehnen 1998), which
hereafter will be referred to as the ``Herculis'' stream according to a
co-moving Eggen group. One of the most reasonable interpretation of this
stream relies on the influence of the Galactic bar, which is believed to have
a corotation radius $R_{\rm CR}\approx 3.5-5$~kpc and a major-axis presently
leading by $\varphi\approx 15^{\circ}-45^{\circ}$ the Sun's azimuth. Raboud et
al. (1998) have shown that $N$-body models of the Milky~Way do produce a mean
outward motion of disc particles at realistic positions for the Sun relative
to the bar and suggest that the Herculis stream involves stars on hot orbits,
i.e. orbits with a high enough value of Jacobi's integral to cross the
corotation radius, whereas Dehnen~(2000, hereafter D2000) relates this
velocity mode and the main low-asymmetric drift mode of the observed $U-V$
distribution to the coexistence near the outer Lindblad resonance (OLR) of the
inner anti-bar and outer bar elongated periodic orbits replacing the circular
orbit close to the OLR in a rotating barred potential (i.e. the same idea
introduced by Kalnajs~1991 to explain the Hyades and Sirius streams), and
attributes the valley at $S\approx 30$~km\,s$^{-1}$ between the two modes
(respectively ``OLR'' and ``LSR'' mode in his terminology) to stars on
unstable OLR orbits. This paper analyses the role of bar induced chaotic
orbits in this context and confronts both alternatives using a similar 
2D modelling technique as in Dehnen's work. The details are presented in
Fux (2000).

\section{Resonances and periodic orbits}

The working potential is the same as in D2000, consisting of the sum of an
axisymmetric logarithmic potential with constant circular velocity $V_{\circ}$
and of a rotating quadrupole term with a bar strength $F=0.15$, where $F$ is
defined as the maximum azimuthal force on the circle passing through the ends
of the bar normalised by the axisymmetric radial force at same radius
($F=8.89\alpha$ in Dehnen's notation). The position in space will be
parametrised by the galactocentric distance relative to the OLR radius
$R_{\rm OLR}=(1+1/\sqrt{2})R_{\rm CR}$ and the angle $\varphi$ relative to the
major axis of bar. A distance of $R/R_{\rm OLR}=1.1$ implies
$R_{\rm CR}=4.26$~kpc if $R=R_{\circ}=8$~kpc.
\par Figure~1 (left) shows the characteristic curves in the Hamiltonian$-x$
plane (at $y=0$ and $\dot{y}>0$) of the main periodic orbits existing near the
OLR and closing after one rotation in the rotating frame, using the labelling
of Contopoulos \& Grosbol (1989). The curves essentially resemble those of the
circular orbit and of the resonant orbits in the axisymmetric limit, except
near the resonances of the circular orbit where bifurcations may transform
into gaps, as is the case at the OLR (e.g. Sellwood \& Wilkinson 1993).
Figure~1 also displays some members of the $x_1(1)$ and $x_1(2)$ families.
The thick orbits are examples of the perpendicularly oriented orbits that
replace the circular orbit near the OLR and which are responsible for the
bimodality in the observed $U-V$ distribution according to D2000. In both
families, the orbits develop loops at high eccentricity, i.e. on the part of
the characteristic curves emerging leftwards from the zero velocity curve and
where $\dot{x}\neq 0$, and the $x_1(1)$ orbits are stable over a remarkable
eccentricity range. Figure 2 shows the location in the $U-V$ plane of the
$-4/1$, $-2/1$ (OLR) and $-1/1$ resonances in the axisymmetric part of the
potential for different space positions, and of the periodic orbits passing
through these positions in the total axisymmetric+bar potential. At
$\varphi=90^{\circ}$ most resonant periodic orbits are stable, except for
instance the $x_1^*(2)$ orbit at $R/R_{\rm OLR}\ga 1.1$, whereas at
$\varphi=0$, most of them are unstable.
\begin{figure}[t!]
\centerline{
\psfig{file=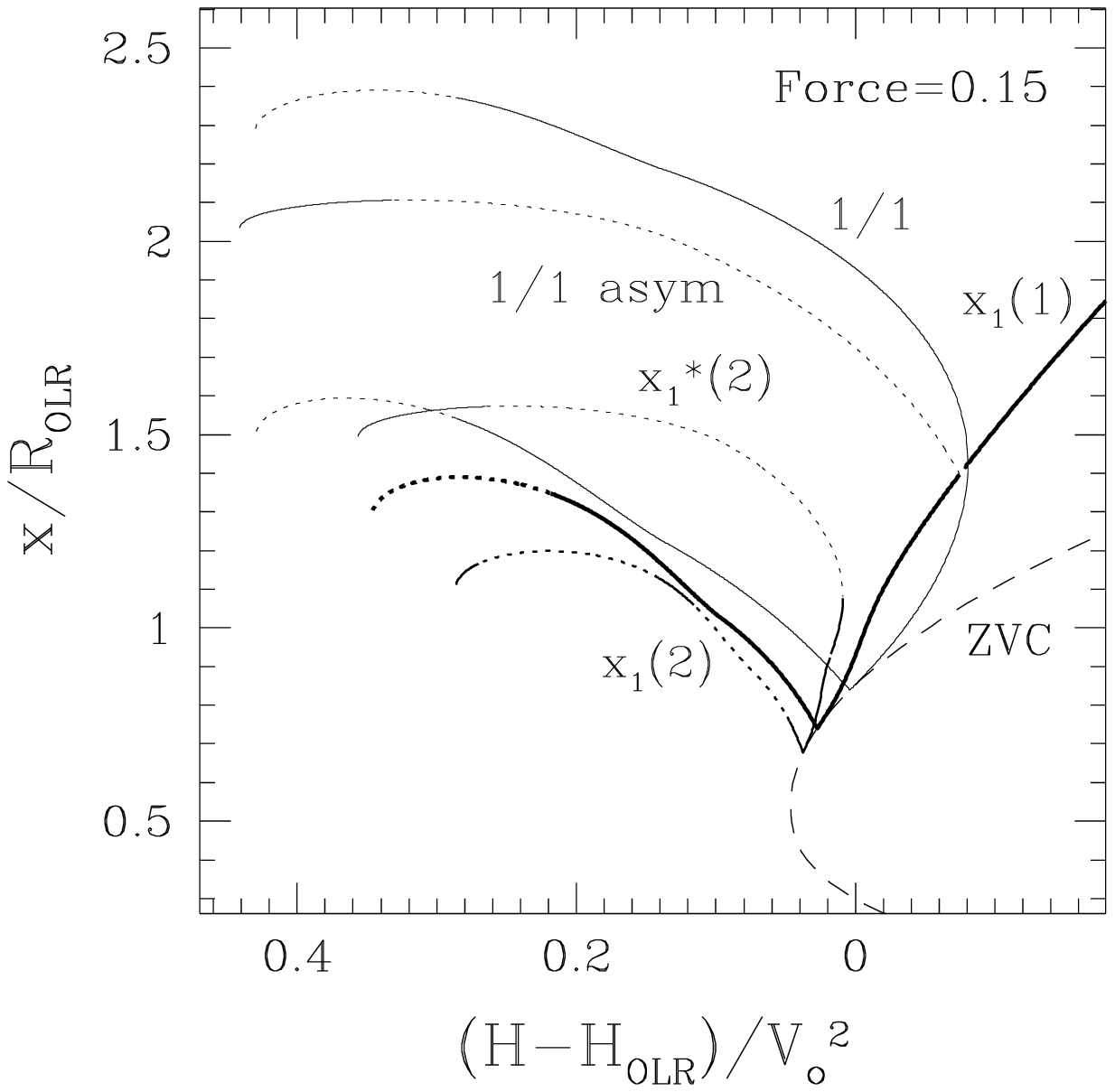,height=4.3cm}\hspace*{.15cm}
\psfig{file=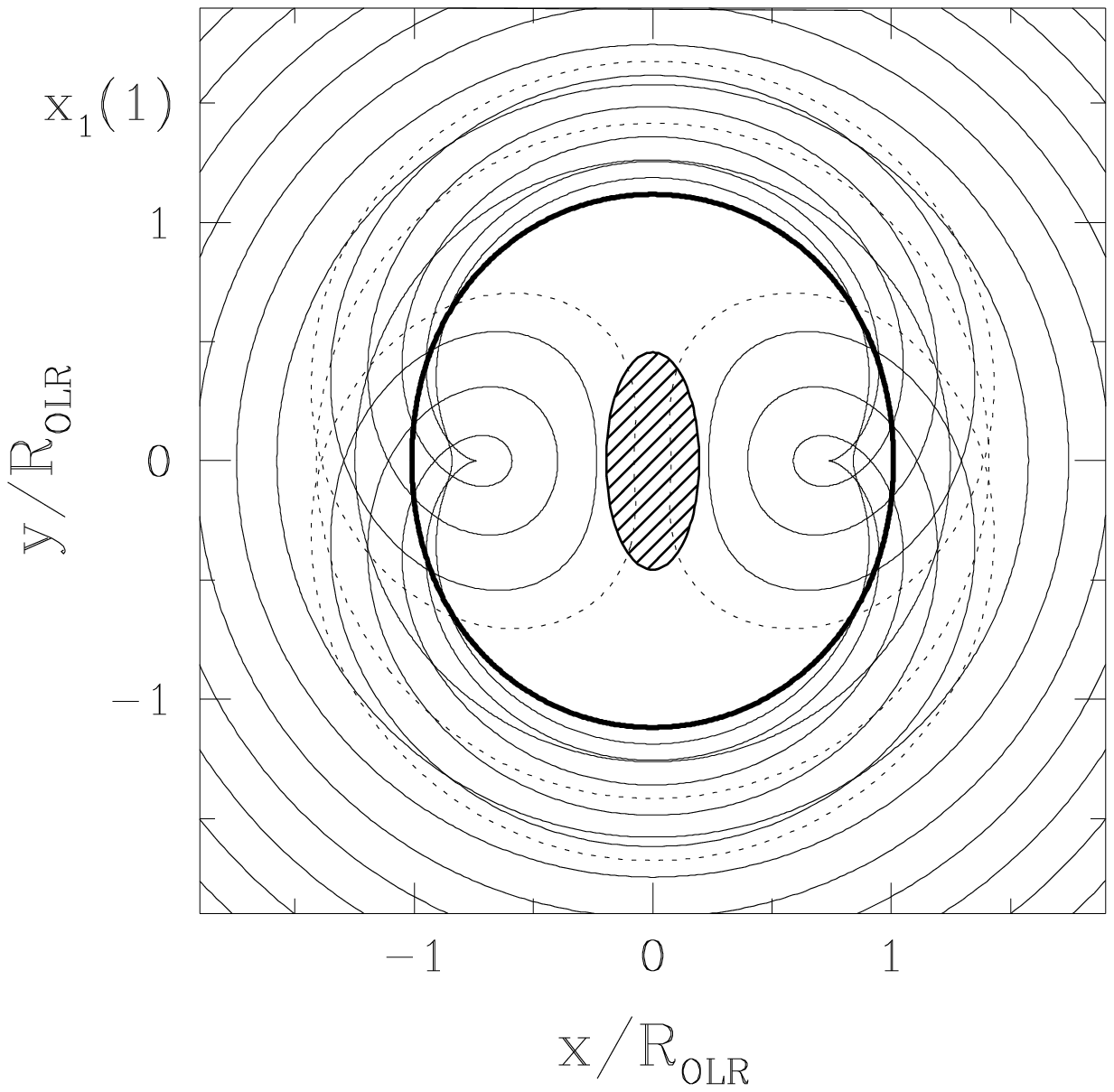,height=4.3cm}\hspace*{-.1cm}
\psfig{file=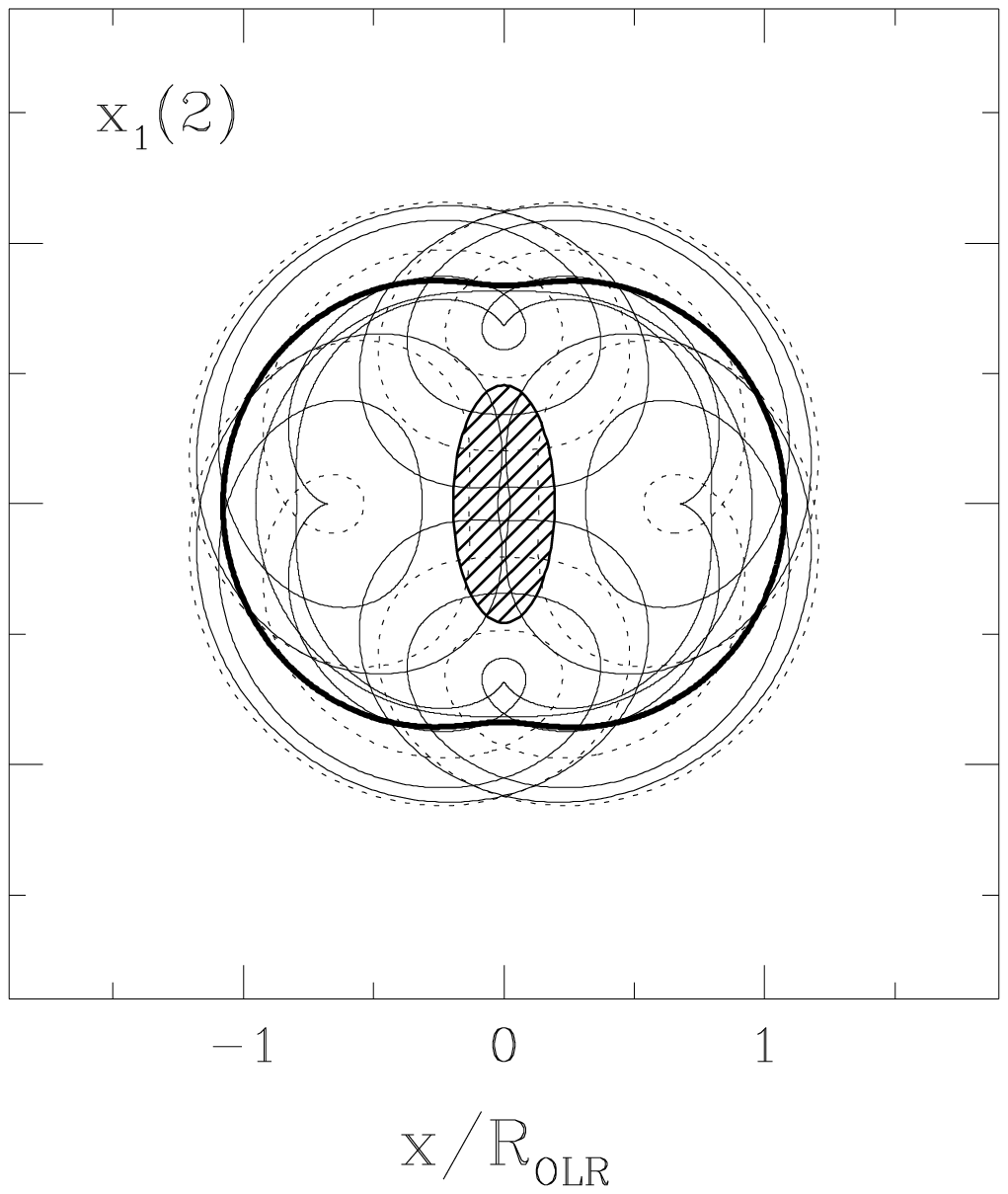,height=4.3cm}}
\caption{Left: characteristic diagram of the relevant periodic orbit families
for $F=0.15$. The dashed line indicates the zero velocity curve and
$H_{\rm OLR}$ is the Hamiltonian of the circular orbit at the OLR in the
axisymmetric part of the potential. Middle and right: sequences of $x_1(1)$
and $x_1(2)$ orbits, with the thick line representing the orbit of lowest
apocentre and of lowest Hamiltonian respectively. The orientation of the bar
is sketched by the shaded ellipse. In all frames, full and dotted lines stand
for stable and unstable orbits respectively.}
\end{figure}

\section{Hot orbits}

In a rotating barred potential, the value of the Hamiltonian
$H=\frac{1}{2}{\dot{\vec{x}}^2}+\Phi_{\rm eff}(\vec{x})$, where
$\Phi_{\rm eff}(\vec{x})$ is the effective potential, is the only classical
integral of motion, known as Jacobi's integral. Two critical values of this
integral are related to the Lagrangian points $L_{1/2}$ and $L_{4/5}$, i.e.
respectively the saddle points and maxima of $\Phi_{\rm eff}(\vec{x})$.
If $H>H_{12}\equiv \Phi_{\rm eff}(L_{1/2})$, an orbit is susceptible to cross
the corotation radius, defining the {\it hot orbit} category, and if
$H>H_{45}\equiv \Phi_{\rm eff}(L_{4/5})$, an orbit is no longer spatially
confined by this integral.
\par If $U$ and $V$ are measured relative to the galactic centre, the contours
of constant $H$ in the $U-V$ plane are circles centred on
$(V,U)=(R\,\Omega_{\rm P},0)$, where $\Omega_{\rm P}$ is the pattern speed
of the bar, and of radius $\sqrt{2(H-\Phi_{\rm eff})}$. Figure~2 gives the
$H_{12}$ and $H_{45}$ contours as a function of space position for the working
potential. The hot orbits are those outside (low $V$) the former contour. At
$R=R_{\rm OLR}$ and $\varphi=30^{\circ}$, the $H_{12}$ contour crosses the
$V$-axis at $S=V_{\circ}-V\approx 0.177V_{\circ}=35$\,km\,s$^{-1}$ for
$V_{\circ}=200$~km\,s$^{-1}$, placing the Herculis stream in the hot orbit
zone.

\section{Regular and chaotic regions of phase-space}

The next step after the periodic and hot orbits is to determine the extension
in phase-space of the regular orbits trapped around the stable periodic
orbits and of the chaotic regions. A convenient tool for this purpose are the
Liapunov exponents, which describe the mean exponential rate of divergence of
two trajectories with infinitesimally different initial conditions. The largest
such exponent, $\lambda_1$, determines the degree of stochasticity of the
orbits and vanishes for regular orbits. Figure 2 gives the divergence
timescale $t_{\lambda}\equiv 1/\lambda_1$ of the orbits as a function of
planar velocity and space position.
\par The Liapunov divergence timescale can be as low as a few orbital times~for
chaotic orbits. As expected, the stable periodic orbits always lie within
regular regions and the unstable orbits within chaotic regions. Furthermore,
the $H_{12}$ contour appears as the average transition from regular motion
at low $H$ to chaotic motion at high $H$, and thus most hot orbits are
chaotic. Due to the four-fold symmetry of the potential, the diagrams at
$\varphi=0$ and $\varphi=90^{\circ}$ are symmetric with respect to $U=0$, and
more generally, diagrams at supplementary angles are anti-symmetric to each
other in $U$. At $\varphi=90^{\circ}$, the $-2/1$ and $-1/1$ resonance curves
are mainly embedded within broad arcs of regular orbits spaced by chaotic
regions at large $H$, while at $\varphi=0$, these regular arcs occur between
the resonance curves. At intermediate angles, the regular and chaotic regions
become offset from the resonance curves and the $U$-symmetry brakes.
In particular, for $\varphi \sim 30^{\circ}$ and for the displayed radial
range, a prominent region of regular orbits trapped around stable eccentric
$x_1(1)$ orbits extends down to negative~$U$, bounded roughly by the OLR curve
on the right and penetrating well inside the hot orbit zone, whereas the
positive $U$ part of the OLR curve is surrounded by a wide chaotic region
extending somewhat inside the $H_{12}$ contour and coinciding very well with
the observed $U-V$ location of the Herculis stream. Note that since 2D
axisymmetric potentials are always integrable, all chaotic regions reported
here owe uniquely to the bar.
\begin{figure}[t!]
\centerline{\psfig{file=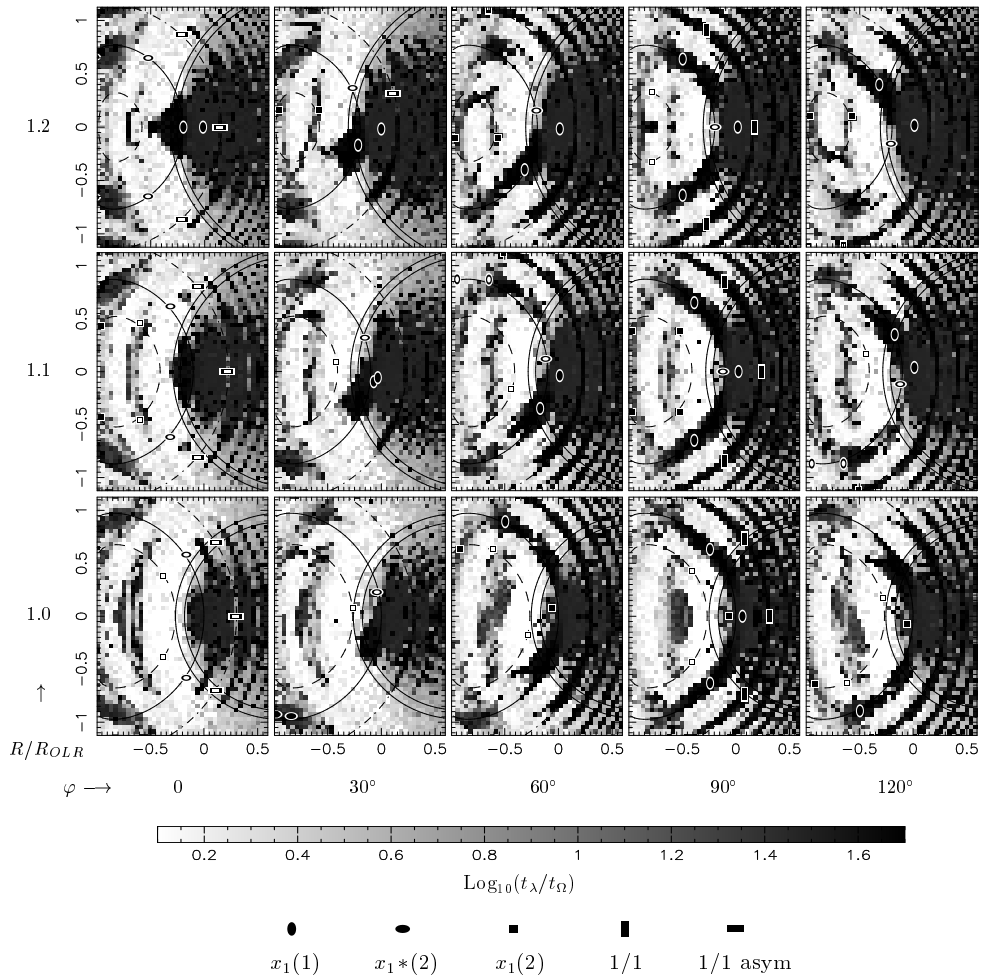,width=13cm}}
\caption{Liapunov divergence timescale of the orbits in the $U-V$ plane
as a function of space position, for $F=0.15$. The horizontal and vertical
axes of each frame are $V/V_{\circ}$ and $U/V_{\circ}$ respectively, with
the origin at the circular orbit of the underlying axisymmetric potential.
The timescale is in units of local circular period $t_{\Omega}$. The dark
and white regions respectively indicate regular and chaotic orbits, the
various full and empty symbols the stable and unstable periodic orbits,
the circular arcs open towards the right the $H_{12}$ and $H_{45}$ contours,
and the dash-dotted, solid and dashed curves the $-1/1$, $-2/1$ and $-4/1$
resonances in the axisymmetric part of the potential.}
\end{figure}
\begin{figure}[t!]
\centerline{\psfig{file=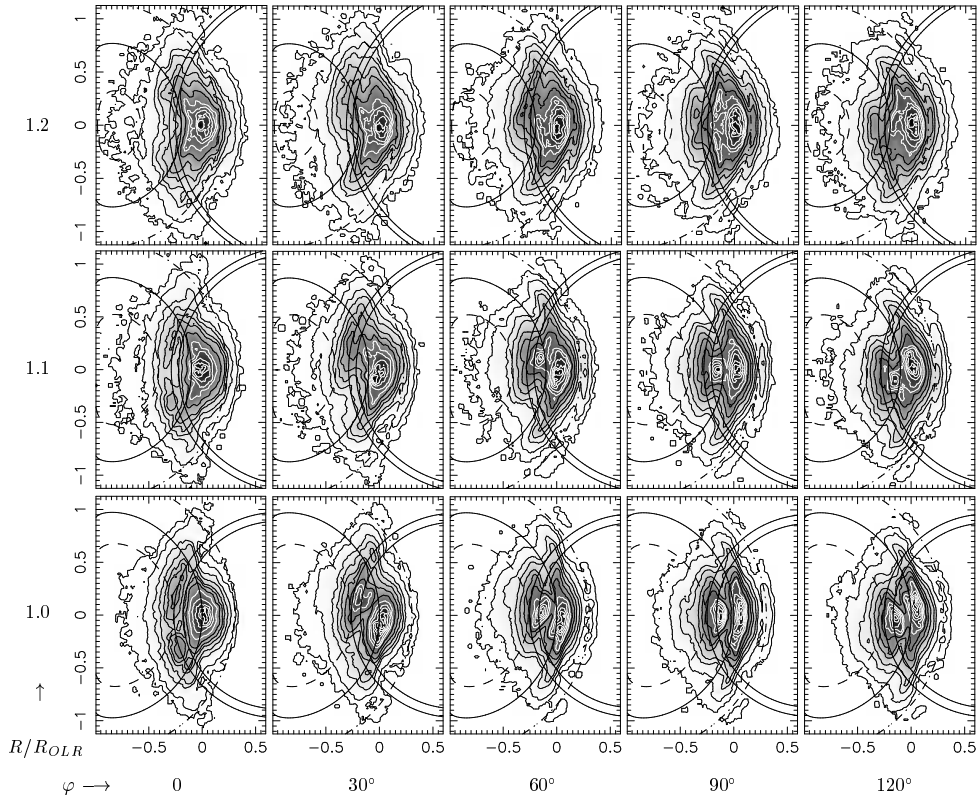,width=13cm}}
\caption{Velocity distribution as a function of bar parameters in the test
particle simulations with $F=0.15$. The distributions are averaged over the
time interval going from $55$ to $65$ bar rotations. The axes of the frames
and the resonance and $H$ contours are as in Figure~2.}
\end{figure}
\par A detailed investigation reveals that the regular orbits associated with
the stable low eccentricity periodic orbits of the $x_1(2)$ family exist for
$R/R_{\rm OLR}\la 1.1$ and over an angle range around $\varphi=90^{\circ}$
increasing as $R$ decreases. Elsewhere, unstable $x_1^*(2)$ orbits have
completely substituted to the regular $x_1(2)$ islands. In particular,
there is no $x_1(2)$ streaming motion expected at $R/R_{\rm OLR}=1.1$ and
$\varphi=30^{\circ}$ in the present working potential.

\section{Phase-space crowding}

Finally, we want to know how nature populates the available orbits. This
is done by test particle simulations starting with exactly the same
initial axisymmetric distribution functions $f_{\circ}$ and growing the bar
the same way (i.e. over $2$ bar rotations) as in Dehnen's default simulations.
However, we do not apply the claimed high-resolution and noise-free backward
integration technique based on the conservation of the phase-space density
along orbits. The reason is because much longer integration times than the few
bar rotations adopted in D2000 are necessary to get phase-mixed and
quasi-steady evolved distribution functions, and that for integration times
larger than about $10$ bar rotations, the fine-grained $U-V$ distributions
resulting from this technique become highly fluctuating on small scales and
even noisy in the chaotic regions and thus must be smoothed to yield the
physical coarse-grained distributions. Therefore, we decided to use the
standard forward integration technique, sampling $f_{\circ}$ with $N=10^6$
particles, and to average the velocity distributions, derived within finite
volumes of radius $R/80$, over $10$ bar rotations to improve the statistics
and also to minimise the phase-mixing problem. The results are depicted in
Figure~3. Here the distance parameters in $f_{\circ}$ always scale as $R$ and
the truly changing parameters are those of the bar. Hence frames at different
$R/R_{\rm OLR}$ come from distinct simulations.
\par The $U-V$ distributions seem to present the same symmetry properties
in~$U$ as described in the previous section for the Liapunov divergence
timescale. This is obviously not the case in D2000, as a consequence of the
too short integration. The high angular momentum peak of the velocity
distributions also coincides better with the trace of the non-resonant
$x_1(1)$ orbits at $U=0$. The regular low eccentricity $x_1(2)$ regions also
produce a peak in the velocity distributions, which always lies inside the
$H_{12}$ contour.
\par In the hot orbit zone, the chaotic regions appear to be more heavily
crowded than the regular regions. This is because the phase-space density
in regular regions remains roughly constant during the simulations, whereas
there is a net migration of particles on hot chaotic orbits from the inner
to the outer space regions, and because the chaotic and regular regions are
decoupled from each other, i.e. chaotic orbits cannot visit the regions
occupied by regular orbits and vice versa. Hence at large $H$, the velocity
distributions are dominated by chaotic orbits which are forced to avoid less
populated regular regions. In~the parameter range considered here, these
forbidden regions are essentially the broad regular orbit regions
around the stable eccentric $x_1(1)$ orbits. In~particular, at
$R/R_{\rm OLR}=1.1$ and $\varphi=30^{\circ}$, this $x_1(1)$ region lies at
$U\la 0$, as mentioned above, and thus the main overdensity of hot chaotic
orbits occurs at positive $U$, providing a natural interpretation of the
Herculis stream. The depression on the right of this stream may be attributed
to the decline of the hot orbit population as $H\rightarrow H_{12}$.
The high eccentricity $x_1(1)$ region is part of Dehnen's valley, but
represents {\it regular} OLR orbits, and the unstable $x_1^*(2)$ orbit does
not always fall within this valley (e.g. at $R/R_{\rm OLR}=1.2$ and
$\varphi=30^{\circ}$).
\par Another property of the hot chaotic regions is that the contours of the
velocity distribution follow the lines of constant Hamiltonian. This can be
understood in terms of the Jeans theorem (see also the answer to Dehnen's
question): in two-dimensions, chaotic orbits have no other integral of motion
than Jacobi's integral and thus the distribution function in their
neighbourhood depends only on this integral, i.e. on the value of the
Hamiltonian.

\clearpage

\section*{Questions}

\subsubsection{Ortwin Gerhard:}
The spiral arms outside the Milky Way's bar region must modify the population
seen in the $U-V$ diagram. Could you say how?
\subsubsection{Roger Fux:}
The effect of spiral arms is not yet very clear. As evolving transient
features, they certainly introduce a time dependency of the $U-V$
distributions and probably increase the amount of chaos. Since Jacobi's
integral is then no longer conserved, they may also introduce some diffusion
from the regular regions to the chaotic regions and vice versa (and the same
remark is true for massive compact objects like giant molecular clouds).
$N$-body simulations displaying spiral arms reveal time averaged $U-V$
distributions that are qualitatively similar to those of the test particle
simulations presented here, but with an azimuthal phase shift apparently
related to the twist of the potential well induced by the spiral arms outside
the bar.

\subsubsection{Walter Dehnen:}
How can stars stay on chaotic orbits, which due to their $H$ are not bound to
the galaxy?
\subsubsection{Roger Fux:}
The timescale for chaotic hot orbits with moderate $H$ to escape the galaxy
is much larger than the orbital timescale and even the Hubble time (see for
instance Kaufmann \& Contopoulos 1996, A\&A 309, 381). In particular, once
phase-mixed, the distribution function in the neighbourhood of such orbits may
be considered as almost steady, which is a necessary condition to apply the
Jeans theorem.

\end{document}